\documentclass[iop,twocolappendix,apj]{emulateapj}

\usepackage{apjfonts,jab,amsmath}
\shortauthors{CHEN ET AL.}
\slugcomment{\today}

\begin{document}

\title{Intermittency of Solar Wind Density Fluctuations From Ion to Electron Scales}
\author{C.~H.~K.~Chen\altaffilmark{1,2}, L.~Sorriso-Valvo\altaffilmark{3,2}, J.~\v Safr\'ankov\'a\altaffilmark{4} and Z.~N\v eme\v cek\altaffilmark{4}}
\affil{$^1$Department of Physics, Imperial College London, London SW7 2AZ, UK; christopher.chen@imperial.ac.uk}
\affil{$^2$Space Sciences Laboratory, University of California, Berkeley, California 94720, USA}
\affil{$^3$IPCF/CNR, UOS di Cosenza, 87036 Rende, Italy}
\affil{$^4$Faculty of Mathematics and Physics, Charles University, Prague 18000, Czech Republic}
\begin{abstract}
The intermittency of density fluctuations in the solar wind at kinetic scales has been examined using high time resolution Faraday cup measurements from the \emph{Spektr-R} spacecraft. It was found that the probability density functions (PDFs) of the fluctuations are highly non-Gaussian over this range, but do not show large changes in shape with scale. These properties are statistically similar to those of the magnetic fluctuations and are important to understanding the dynamics of small scale turbulence in the solar wind. Possible explanations for the behavior of the density and magnetic fluctuations are discussed.
\end{abstract}

\section{Introduction}

The solar wind is thought to contain a turbulent cascade of energy in fluctuations from large, system-size scales, to small, plasma kinetic scales \citep[e.g.,][]{carbone12,horbury12,bruno13,alexandrova13a}. An important feature of this cascade is intermittency, which has traditionally referred to the bursty, non-Gaussian nature of turbulent fluctuations \citep[e.g.,][]{batchelor49}, although more recent definitions specify an increase in the non-Gaussianity towards smaller scales \citep[e.g.,][]{frisch95}.

Intermittency has been extensively measured at magnetohydrodynamic (MHD) scales (larger than the ion gyroradius) in the solar wind \citep[e.g.,][]{
burlaga91a,marsch94,horbury97a,sorriso-valvo99,veltri99,hnat02a,bruno03,kiyani09a,yordanova09,wan12b}, where the probability density functions (PDFs) of fluctuations of various fields are seen to be non-Gaussian. In most cases, they become more non-Gaussian towards smaller scales, and structure function scaling exponents are non-linear, indicative of a multi-fractal cascade.

In the kinetic scale range, it is thought that there is a further cascade of energy from ion to electron scales \citep[e.g.,][]{ghosh96,stawicki01,schekochihin09,boldyrev12b}. The magnetic fluctuations here are measured to be non-Gaussian, although different studies have reported the amount of non-Gaussianity to either increase or remain the same towards smaller scales \citep{alexandrova08b,kiyani09a,kiyani13,wu13}. Characterizing this intermittency is important for understanding the distribution of energy in the kinetic scale cascade \citep{boldyrev12b} and how it is dissipated at electron scales \citep{wan12a,tenbarge13a}. Recently, a model of the effect of intermittency on the energy spectrum of strong kinetic Alfv\'en turbulence was proposed \citep{boldyrev12b}, in which the development of 2D sheets leads to a perpendicular wavenumber spectrum $\propto k_\perp^{-8/3}$, rather than $\propto k_\perp^{-7/3}$ for the non-intermittent case \citep[e.g.,][]{vainshtein73,biskamp99a,cho09a,schekochihin09}.

While density fluctuations have long been measured at MHD scales \citep[e.g.,][]{celnikier83}, they have only recently been measured in the kinetic range. Their spectrum was shown to match that of the magnetic field \citep{chen12a,safrankova13a} and their amplitude, relative to the magnetic fluctuations, was used to infer the predominantly kinetic Alfv\'en, rather than whistler, nature of the turbulence \citep{chen13c}. At MHD scales, anisotropy is thought to lead to density fluctuations being passive to the Alfv\'enic turbulence \citep{goldreich95,lithwick01,schekochihin09}, but in the kinetic range they are likely to be an active component of the kinetic Alfv\'en turbulence \citep{schekochihin09,chandran09c,boldyrev12b,boldyrev13a}, on an equal footing with the magnetic fluctuations. As far as we are aware, there have been no previous measurements of the intermittency of density fluctuations in this range. In this Letter, we present such measurements, compare them to the magnetic fluctuations and discuss the implications for our understanding of kinetic scale turbulence.

\section{Data Set}

To measure density fluctuations between ion and electron scales, a high time resolution data set is required. One possibility is the spacecraft potential measurement from the \emph{ARTEMIS} spacecraft, which has been used to measure the density fluctuation spectrum in this range \citep{chen12a,chen13a,chen13c}. However, large amplitude harmonics of the spacecraft spin frequency (due mainly to the varying photoelectron emission) make time domain analysis difficult.

\begin{deluxetable*}{ccccccccc}
\tablecaption{\label{tab:intervals}List of interval parameters. Data are from \emph{Spektr-R}, except $B$ and $T_\mathrm{e}$, which are from the upstream \emph{Wind} spacecraft.}
\tablehead{\colhead{Interval} & \colhead{Date (dd/mm/yyyy)} & \colhead{Time (UT)} & \colhead{$B$ (nT)} & \colhead{$n_\mathrm{i}$ (cm$^{-3}$)} & \colhead{$v_\mathrm{i}$ (km s$^{-1}$)} & \colhead{$T_\mathrm{i}$ (eV)} & \colhead{$T_\mathrm{e}$ (eV)} & \colhead{$\beta_\mathrm{i}$}}
\startdata
1 & 10/11/2011 & 15:30:00--19:00:00 & 4.6 & 4.7 & 370 & 9.6 & 11 & 0.87\\
2 & 23/04/2012 & 08:30:00--10:20:00 & 10 & 25 & 370 & 14 & 10 & 1.3\\
3 & 01/06/2012 & 21:10:00--26:09:00 & 8.5 & 6.6 & 370 & 3.2 & 6.6 & 0.12\\
4 & 09/07/2012 & 11:20:00--14:10:00 & 12 & 5.4 & 390 & 2.6 & 13 & 0.04\\
5 & 09/08/2012 & 09:10:00--22:40:00 & 4.6 & 5.1 & 320 & 7.4 & 13 & 0.73\\
6 & 13/11/2012 & 06:10:00--09:00:00 & 11 & 20 & 440 & 25 & 13 & 1.8
\enddata
\end{deluxetable*}

In this Letter, data from the BMSW instrument \citep{safrankova13b} of the Plasma-F experiment on the \emph{Spektr-R} spacecraft were used. BMSW has six Faraday cups, which, together, can sample the solar wind ion distribution at 32 samples/s. Three of the Faraday cups face Sunward and continuously measure the current of incoming ions above given energy thresholds set by positive voltages on their control grids. When in adaptive mode, the voltages on two of these Faraday cups are varied by feed-back loops so that fixed fractions of the total ion current, determined by the third, are measured. The other three are inclined by 20$^\circ$ and used mainly to determine the solar wind direction. Assuming an isotropic Maxwellian ion distribution, these six data points can be used to infer the ion density $n_\mathrm{i}$, velocity $\mathbf{v}_\mathrm{i}$ and temperature $T_\mathrm{i}$ every 31.25 ms. While an isotropic Maxwellian is not always appropriate for the solar wind, data from SWE on the upstream \emph{Wind} spacecraft were checked to ensure that intervals used in this analysis had approximately isotropic proton temperatures $T_\perp\approx T_\|$.

To determine the kinetic scales and ion plasma beta $\beta_\mathrm{i}$, the magnetic field $\mathbf{B}$ is required. Since this is not available from \emph{Spektr-R}, data from MFI on \emph{Wind}, lagged by eye to provide the best match between density, velocity and temperature features in BMSW and \emph{Wind} 3DP, were used.

\section{Results}

We begin by presenting the results of the analysis for one 3.5 hr interval of data (Interval 1, Table \ref{tab:intervals}). Since neither the full ion distribution, nor the electromagnetic fluctuations were measured, it is not possible to determine with certainty whether the spacecraft was in the foreshock, but the density and velocity power spectra are not dominated by the signatures of foreshock waves \citep[e.g.,][]{fazakerley95,eastwood02}.

\begin{figure}
\includegraphics[width=\columnwidth]{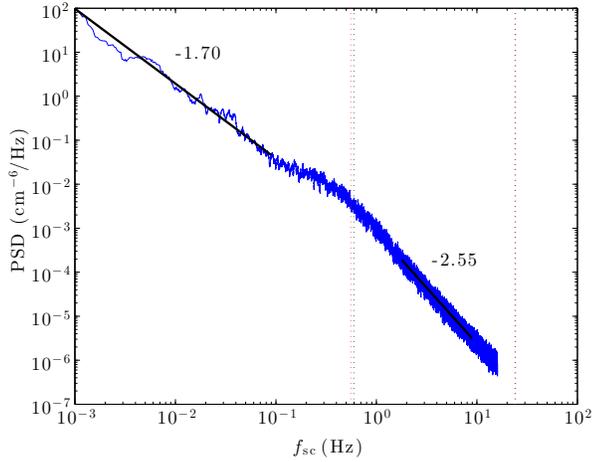}
\caption{\label{fig:spectrum}Spectrum of ion density fluctuations in Interval 1. Spacecraft-frame frequencies $f_\mathrm{sc}$ corresponding to the proton gyroscale (left black dotted), electron gyroscale (right black dotted), proton inertial length (left red dotted) and electron inertial length (right red dotted) are marked.}
\end{figure}

The power spectrum of ion density fluctuations, as a function of spacecraft-frame frequency $f_\mathrm{sc}$, is shown in Figure \ref{fig:spectrum}. Frequencies corresponding to the proton and electron gyroradii and inertial lengths have been marked, assuming the Taylor hypothesis, along with the power law fits $\alpha=-1.70$ and $\alpha=-2.55$ in each range. It can be seen that the spectrum has a similar shape to that of the electron density fluctuations \citep{chen12a,chen13c}, including the ion scale flattening \citep[e.g.,][]{celnikier83,chen13a}, as expected due to quasi-neutrality.

\begin{figure}
\includegraphics[width=\columnwidth]{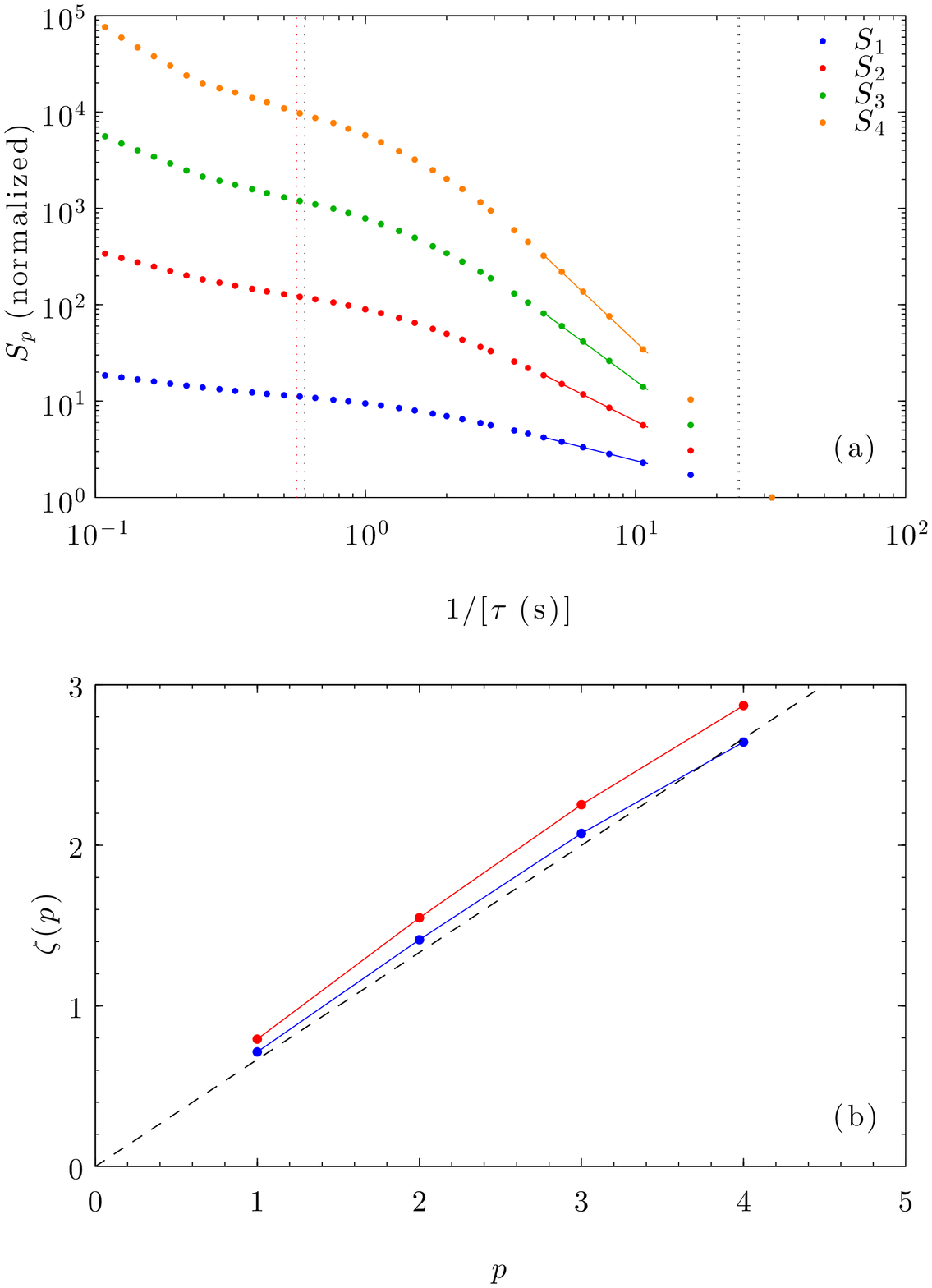}
\caption{\label{fig:sfun}(a) Structure functions $S_p$ as a function of inverse scale $1/\tau$ normalized to their $\tau=31.25$ ms values for clarity. Power law fits are shown as solid lines. Kinetic scales are marked as in Figure \ref{fig:spectrum}. (b) Structure function scaling $\zeta(p)$ calculated directly from the structure functions (blue) and using ESS (red). The dashed line is the non-intermittent scaling $\zeta(p)=\frac{2}{3}p$.}
\end{figure}

One way to examine the intermittency of these fluctuations is from the scaling properties of their structure functions. These can be defined as
\begin{equation}
S_p(\tau)=\left<|\delta n_\mathrm{i}(t,\tau)|^p\right>,
\end{equation}
where $\delta n_\mathrm{i}(t,\tau)=n_\mathrm{i}(t+\tau)-n_\mathrm{i}(t)$, $\tau$ is a time lag, and the angular brackets denote an average over times $t$, and they typically take a power law form $S_p(\tau)\propto\tau^{\zeta(p)}$ in the inertial range of a turbulent cascade. Figure \ref{fig:sfun}a shows the first four structure functions for time lags from ion to electron scales, plotted as a function of $1/\tau$ for comparison to the spectrum. A rule of thumb for the maximum order that can be determined reliably from a sample of $N$ points is $p_\mathrm{max}=\log N-1$ \citep{dudokdewit13}, giving $p_\mathrm{max}=4$ for $N=396,517$ in this interval. A more rigorous estimate, using the technique of \citet{dudokdewit04}, gives $p_\mathrm{max}=3$. It can be seen that the structure functions form approximate power laws but only in the smaller scale part of the range, due to the ion scale flattening ($\tau\approx$ 0.5 s to 5 s), which is more prominent here than in the spectrum.

Figure \ref{fig:sfun}b shows, in blue, the power law fits $\zeta (p)$ for scales $\tau=$ 94 ms to 220 ms, the approximately power law range. Error bars from the fits are smaller than the data points, although the scaling may be affected by the curvature of the structure functions. Extended Self-Similarity (ESS) is an technique to obtain the scaling, which can be used to overcome such curvature \citep{benzi93}. Since $S_p\propto S_2^{\zeta(p)/\zeta(2)}$, $S_p$ can be plotted as a function of $S_2$ and the slope used to determine $\zeta(p)/\zeta(2)$. $\zeta(2)$ can then be obtained from the spectral index in Figure \ref{fig:spectrum}, $\zeta(2)=-(1+\alpha)$, to give $\zeta(p)$. This technique was used to obtain the scaling exponents over a larger range, from $\tau=$ 94 ms to 1 s, to match the --2.55 power law range in Figure \ref{fig:spectrum}, and they are shown in Figure \ref{fig:sfun}b in red. Again, the errors from the fits are smaller than the data points. It can be seen that the ESS exponents are steeper than the non-intermittent prediction of $\zeta(p)=\frac{2}{3}p$ (corresponding to a $k^{-7/3}$ spectrum) and are almost linear. The near linearity suggests a mono-fractal, or perhaps weakly multi-fractal cascade, although with reliable exponents up to $p_\mathrm{max}=3$, it is not easy to make this distinction.

\begin{figure}
\includegraphics[width=\columnwidth]{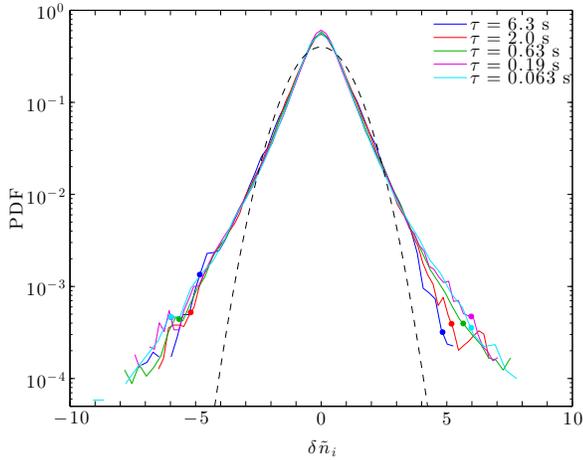}
\caption{\label{fig:pdfs}Probability density functions (PDFs) of normalized ion density fluctuations $\delta\tilde{n}_{\text{i}}$ at scales $\tau$ from ion to electron scales, with a Gaussian of unit variance for comparison (black dashed).}
\end{figure}

It is perhaps more instructive to examine the full PDFs of the ion density fluctuations to determine the scale dependence of their non-Gaussianity. Figure \ref{fig:pdfs} shows the PDFs of the normalized fluctuations 
\begin{equation}
\label{eq:dn}
\delta\tilde{n}_\mathrm{i}(t,\tau)=\frac{\delta n_\mathrm{i}(t,\tau)}{\sqrt{\left<\delta n_\mathrm{i}(t,\tau)^2\right>}},
\end{equation}
from ion to electron scales, where clear non-Gaussian behavior can be seen. The minimum bin size in the PDF of 0.01 cm$^{-3}$ is comparable to the instrumental noise level, and bins with fewer than 10 points were removed, so that the Poisson errors on the PDF are $<$ 32\%. Although there is a slight tendency of becoming more non-Gaussian towards smaller scales, the PDFs remain approximately similar in shape over this range, which covers both the flattening and power law ranges. The PDFs of ion velocity and temperature fluctuations were also examined, but were found to become Gaussian at electron scales, consistent with their domination by amplifier noise here \citep{safrankova13b}.

\begin{figure}
\includegraphics[width=\columnwidth]{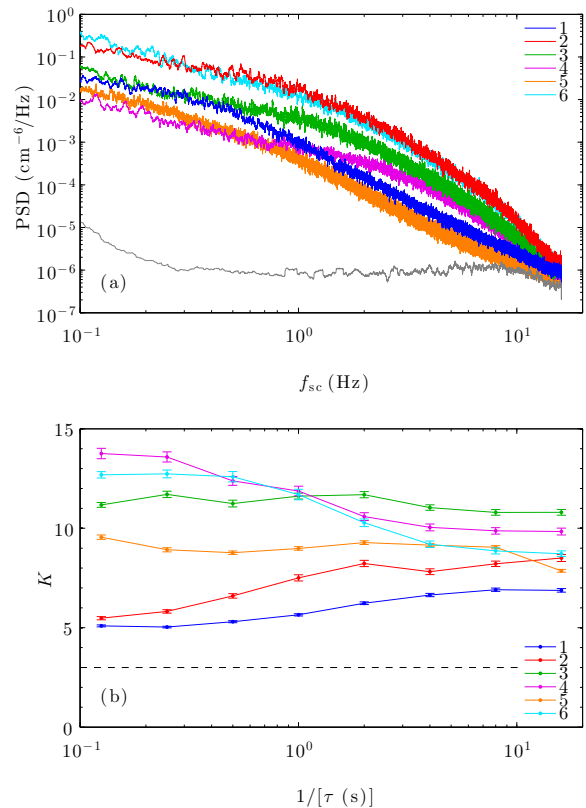}
\caption{\label{fig:kurtosis}(a) Power spectra of ion density fluctuations for all six intervals. The noise spectrum is also shown (grey). (b) Kurtosis $K$ of ion density fluctuations from ion to electron scales as a function of inverse scale $1/\tau$ for all six intervals. The value for a Gaussian ($K=3$) is marked (black dashed).}
\end{figure}

The degree of non-Gaussianity in the density fluctuations can be quantified using the kurtosis, also known as flatness \citep[e.g.,][]{frisch95}, defined as
\begin{equation}
K(\tau)=\frac{\left<\delta n_\mathrm{i}(t,\tau)^4\right>}{\left<\delta n_\mathrm{i}(t,\tau)^2\right>^2}.
\end{equation}
For a Gaussian PDF $K=3$, and for a more peaked PDF with heavier tails $K>3$. Since the maximum reliable order for this interval was determined to be $p_\mathrm{max}=3$, and the kurtosis is based on the 4th moment, it is necessary to investigate its reliability. Errors in calculating $K$ can originate from the large amplitude fluctuations in the tails of the PDFs, which may not be sufficiently sampled, such that the integral in the moment calculation is not well estimated \citep[e.g.,][]{dudokdewit04}. Various schemes exist to deal with this, mostly involving the removal of the largest fluctuations to recover a more reliable scaling. Here, a scheme based on that of \citet{kiyani06} was employed, in which a small fraction of them was removed until the moments appeared converged. Similarly to that study, it was found that removing the largest 0.1\% of fluctuations was sufficient, and the resulting clipping points are marked as dots on the PDFs in Figure \ref{fig:pdfs}. This conditioning was not found to change the qualitative properties of the kurtosis, but removed a few spurious values caused by the finite data length.

The scale-dependent kurtosis was determined for all six intervals in Table \ref{tab:intervals}. The power spectra for these intervals are shown in Figure \ref{fig:kurtosis}a, along with an estimate of the noise spectrum\footnote{This spectrum was determined using data from an in-flight calibration interval in which a high voltage was applied to the control grids of the Faraday cups to exclude the majority of the solar wind ions. The resulting spectrum reflects the amplifier noise, although its precise level will vary slightly with solar wind speed and temperature.}. It can be seen that in the majority of this range the fluctuations are well above the noise, although begin to become affected at the very highest frequencies. The kurtosis for the intervals is shown in Figure \ref{fig:kurtosis}b, where the error bars represent 2 standard errors, determined by the Monte Carlo bootstrap method. Only relatively small changes with scale can be seen: the difference in $K$ between ion and electron scales is less than a factor of 2 in all cases. Intervals with smaller $K$ at ion scales show a slight increase towards smaller scales and vice versa, features consistent with the PDF shapes. The two intervals with an increasing kurtosis (1 and 2) are within a few $R_\mathrm{E}$ the bow shock, so while there are no conclusive signatures of foreshock waves, it is possible that a low level of such activity could be causing the smaller values of $K$ at ion scales. Intervals 3-5, which are 25--40 $R_\mathrm{E}$ upstream of Earth, are unlikely to be affected in this way. Interval 6 is also within a few $R_\mathrm{E}$ of the bow shock, but shows a decrease in $K$ towards smaller scales. Where the instrumental noise becomes significant ($\tau\lesssim$ 0.14 s), the values of $K$ may be unreliable, for example the decrease in Interval 5 is likely unphysical, but this affects only the smallest scales. In general, while the behaviour of the kurtosis is somewhat variable between intervals, the changes with scale are not large.

\section{Discussion}

We have shown that the PDFs of ion density fluctuations do not display large changes in shape from ion to electron scales, with almost linear structure function scaling exponents. This appears to be the case over the majority of the range, although instrumental noise makes interpretation of the intermittency close to electron scales difficult. This behavior is similar to previous measurements of the magnetic fluctuations in this range \citep{kiyani09a,kiyani13,wu13}, although is different to the large increase in kurtosis towards electron scales reported by \citet{alexandrova08b}. To investigate this difference, the same analysis was performed on the magnetic fluctuations \citep[using the interval of][]{chen10b} and the PDFs were also seen to remain similar in shape. It appears, therefore, that both the density and magnetic fluctuations have similar statistical properties in the kinetic scale range.

There are several possibilities for this behavior of the density and magnetic fluctuations. One possibility is that kinetic scale turbulence is predominantly mono-fractal in nature. This difference to the MHD range is plausible, since the non-linear terms in the relevant fluid equations are different \citep{schekochihin09,boldyrev13a}. Mono-fractal scaling was recently reported in reduced Hall MHD simulations \citep{rodriguezimazio13,martin13}, although earlier electron MHD simulations showed the PDFs becoming more non-Gaussian towards smaller scales \citep{cho09a}. In both cases, however, the PDFs were close to Gaussian at ion scales, whereas in the solar wind they are already strongly non-Gaussian.

An alternative explanation is that the kurtosis is limited by another process. In hydrodynamic turbulence, the kurtosis is seen to saturate at the dissipative scales, after a rapid increase \citep{chevillard05}. It has been argued that the kurtosis of MHD turbulence should saturate at small scales if the dissipative structures are current sheets, since above a given aspect ratio they would become unstable \citep{biskamp90}. The limited kurtosis increase could also reflect the effect of damping, which has been suggested to cause the turbulence to become weaker here \citep{howes11c}. Finally, the presence of incoherent waves was suggested to have caused the kurtosis of magnetic fluctuations to decrease at ion scales \citep{wu13}.

The fact that the density and magnetic fluctuations have similar non-Gaussian statistical properties is consistent with the arguments of \citet{boldyrev12b} that the non-linear terms cause the flux function to become striated along density gradients, while the linear terms cause the density fluctuations to equalize with the magnetic fluctuations. We have not, however, been able to test the direct relationship between density and magnetic structures, since magnetic field data is not available from \emph{Spektr-R}. For future studies, and to confirm the results suggested here, longer data sets would be desirable, with density and magnetic fluctuations (and other fields) measured simultaneously at high time resolution, with minimal noise and interference.

\acknowledgments
C. H. K. Chen is supported by an Imperial College Junior Research Fellowship. This work was also supported by NASA contracts NNN06AA01C and NAS5-02099. J.~\v Safr\'ankov\'a and Z.~N\v eme\v cek are supported by GACR contract P209/12/1774. C. H. K. Chen and L. Sorriso-Valvo acknowledge the Marie Curie Project FP7 PIRSES-2010-269297 -- ``Turboplasmas''. \emph{Wind} data were obtained from SPDF ({http://spdf.gsfc.nasa.gov}). We thank S. Boldyrev, J. P. Eastwood, T. S. Horbury, K. H. Kiyani, and A. A. Schekochihin for useful discussions.

\bibliographystyle{apj}
\bibliography{bibliography}

\end{document}